\documentclass[usenatbib]{mn2e}
\pdfoutput=1
\usepackage{graphicx}
\usepackage{amsmath}
\usepackage{placeins}
\usepackage{mathptmx}
\usepackage{txfonts}
\usepackage[T1]{fontenc}
\usepackage{ae,aecompl}
\usepackage{placeins}
\usepackage{hyperref}
\usepackage{pdfpages}
\usepackage{multicol}
\usepackage[english]{babel}
\usepackage{multirow}
\usepackage{graphics}
\usepackage{colortbl}

\usepackage[caption=false]{subfig}
\usepackage{varioref}
\hypersetup{colorlinks,
citecolor=blue
}

\def\gr{general relativity}

\labelformat{section}{Section #1} 
\labelformat{subsection}{Section #1} 
\labelformat{subsubsection}{Section #1}
\labelformat{subsubsubsection}{Section #1}
\labelformat{equation}{Eq.~(#1)} 
\labelformat{figure}{Fig.~#1} 
\labelformat{subfigure}{Fig.~\thefigure#1} 
\labelformat{table}{Table~#1} 
\labelformat{appendix}{Appendix #1}

\title{Implications of Einstein-Maxwell dilaton-axion gravity from the black hole continuum spectrum}
\author[Banerjee, Mandal \& Sengupta]
       {Indrani Banerjee$^{1}$\thanks{E-mail :tpib@iacs.res.in},
       Bhaswati Mandal$^1$\thanks{E-mail : tpbm3@tpbm3.ac.in}
       and Soumitra Sengupta$^{1}$\thanks{E-mail :tpssg@iacs.res.in}  \\
$^1$  School of Physical Sciences, Indian Association for the Cultivation of Science,2A \& 2B Raja S. C. Mullick Road, Kolkata-700032, India
}

\date{ } 
\begin{document}
\maketitle 

\begin{abstract}
String inspired models can serve as potential candidates to replace general relativity (GR) in the high energy/high curvature regime where quantum gravity is expected to play a vital role. Such models not only subsume the ultraviolet nature of gravity but also exhibit promising prospects in resolving issues like dark matter and dark energy, which cannot be adequately addressed within the framework of GR. The Einstein-Maxwell dilaton-axion (EMDA) theory, which is central to this work is one such string inspired model arising in the low energy effective action of the heterotic string theory with interesting implications in inflationary cosmology and in the late time acceleration of the universe. It is therefore important to survey the role of such a theory in explaining astrophysical observations, e.g. the continuum spectrum of black holes which is expected to hold a wealth of information regarding the background metric. The Kerr-Sen spacetime corresponds to the exact, stationary, and axisymmetric black hole solution in EMDA gravity, possessing dilatonic charge and angular momentum originating from the axionic field. In this work, we compute the theoretical spectrum from the accretion disk around quasars in the Kerr-Sen background assuming the thin accretion disk model due to Novikov \& Thorne. This is then used to evaluate the theoretical estimates of optical luminosity for a sample of eighty Palomar-Green quasars which are subsequently compared with the available observations. Our results based on $\chi^2$ analysis indicate that the dilaton parameter $r_2\sim 0.2$ is favored by optical observations of quasars which is further corroborated by other error estimators e.g., the Nash-Sutcliffe efficiency, the index of agreement and their modified versions. We further report that strong dilaton charges ($r_2>1.6$) are disfavored by quasar optical data and the spins associated with the quasars are also estimated.

\end{abstract}


\section{Introduction}
\label{Intro}
The remarkable agreement of \gr\ (GR) with a host of experimental tests \citep{Will:2005yc,Will:1993ns,Will:2005va,Berti:2015itd} makes it the most successful theory of gravity, till date. With the advent of advanced ground-based and space-based missions, the predictions of GR, e.g. presence of black holes and gravitational waves \citep{Abbott:2017vtc,TheLIGOScientific:2016pea,Abbott:2016nmj,TheLIGOScientific:2016src,Abbott:2016blz}, have received ever-increasing observational confirmations. Yet, the quest for a more complete theory of gravity continues, as GR is marred with the black hole and big-bang singularities \citep{Penrose:1964wq,Hawking:1976ra,Christodoulou:1991yfa} and the quantum nature of gravity continues to be elusive \citep{Rovelli:1996dv,Dowker:2005tz,Ashtekar:2006rx}. On the observational front, \gr\ falls short in resolving the nature of dark matter and dark energy \citep{Bekenstein:1984tv,Perlmutter:1998np,Riess:1998cb}, often invoked to explain the galactic rotation curves and the accelerated expansion of the universe, respectively.   

This has given birth to a wide variety of alternative gravity models e.g. higher curvature gravity \citep{Nojiri:2003ft,Nojiri:2006gh,Capozziello:2006dj,Lanczos:1932zz,Lanczos:1938sf,Lovelock:1971yv,Padmanabhan:2013xyr}, extra-dimensional models \citep{Shiromizu:1999wj,Dadhich:2000am,Harko:2004ui,Carames:2012gr} and the scalar-tensor/scalar-vector-tensor theories of gravity \citep{Horndeski:1974wa,Sotiriou:2013qea,Babichev:2016rlq,Charmousis:2015txa} which can potentially fulfill the deficiencies in GR, yielding GR in the low energy limit. 
Among the various alternatives to GR, string theory provides an interesting theoretical framework for quantum gravity and force unification. It is often said that it is difficult to detect any signature of string theory in low energy regime.
In this context it is important to note that string theory in itself is not a model but provides a framework for building string inspired 4D models which can be confronted with the available observations \citep{Cicoli:2020bao}. 

This raises the question whether we can discover some footprints of a stringy model in low energy observations. With this aim in mind, in this work we explore the observational signatures of the Einstein-Maxwell dilaton-axion (EMDA) gravity which arises in the low energy effective action of superstring theories \citep{Sen:1992ua} when the ten dimensional heterotic string theory is compactified on a six dimensional torus $T^6$. Such a scenario consists of $N=4$, $d=4$ supergravity coupled to $N=4$ super Yang-Mills theory and can be appropriately truncated to a pure supergravity theory exhibiting $S$ and $T$ dualities. The bosonic sector of this $N=4$, $d=4$ supergravity coupled to the $U(1)$ gauge field is known as the Einstein-Maxwell dilaton-axion (EMDA) gravity \citep{Rogatko:2002qe} which provides a simple framework to study classical solutions.
Such a theory comprises of the scalar field dilaton and the pseudo scalar axion coupled to the metric and the Maxwell field. The dilaton and the axion fields are inherited from string compactifications and have interesting implications in the late time acceleration of the universe and inflationary cosmology \citep{Sonner:2006yn,Catena:2007jf}. It is therefore worthwhile to explore the role of such a theory in astrophysical observations. This requires one to look for black hole solutions of these string inspired low-energy effective theories. Fortunately, there exists various classes of black hole solutions bearing non-trivial charges associated with the dilaton and the anti-symmetric tensor gauge fields \citep{Gibbons:1987ps,Garfinkle:1990qj,Horowitz:1991cd,Kallosh:1993yg}. The stationary and axi-symmetric black hole solution in EMDA gravity corresponds to the charged, rotating Kerr-Sen metric \citep{Sen:1992ua} where the electric charge stems 
from the axion-photon coupling and not the in-falling charged particles. Also, the axionic field renders angular momentum to such black holes. Investigating the impact of such a background on the available electromagnetic observations is important since this not only provides a testbed for string theory but also a scope to verify the Kerr hypothesis \citep{Johannsen:2016uoh,Bambi:2015kza,Krawczynski:2018fnw}.

This has been explored extensively in the past \citep{Gyulchev:2006zg,An:2017hby,Younsi:2016azx,Hioki:2008zw,Narang:2020bgo} in the context of null geodesics, strong gravitational lensing and black hole shadow. \cite{Gyulchev:2006zg} studied the differences in properties of light propagation in the vicinity of the equatorial plane of the Kerr and the Kerr-Sen background by numerically computing the values of the stong lensing parameters around the supermassive black hole Sgr A* at the centre of our galaxy. They report significant differences in the observable strong lensing parameters depending on the background metric and discuss the scope of accomodating or falsifying the presence of dilaton-axion black holes with the advent of very long baseline interferometry high
resolution imaging in future. \cite{Younsi:2016azx} developed a new parametrization to represent any general stationary and axi-symmetric black hole metric and used it for performing ray-tracing calculations to compute the structure of the black hole shadow in a given stationary and axi-symmetric background. They showed that with the parametrized formulation of the metric, accurate computation of the black hole shadow is possible in the context of Kerr-Sen, Johannsen-Psaltis and Einstein-dilaton-Gauss-Bonnet spacetimes which is expected to gain relevance with the availability of VLBI observations of supermassive black holes like Sgr A*. \cite{Hioki:2008zw} studied null geodesics, circular photon orbits and hidden symmetries in the Kerr-Sen background and discussed the presence of Killing tensors which renders separability of the Hamilton-Jacobi equations in the Kerr-Sen background. They studied the differences in the properties of the capture region and the scattering of photons in the Kerr-Sen and the Kerr-Newman spacetimes and discussed their importance in the context of future astrophysical observations. \cite{An:2017hby} explored the properties of collision (e.g. the centre of mass energy) of uncharged, spinning particles in the Kerr-Sen background and discussed the differences with that of the Kerr and the Kerr-Newman spacetimes. However, none of the above works have reported any constraint on the parameters of the Kerr-Sen metric from the present available observations.

Since deviation from Einstein gravity is expected in the high curvature domain, the near horizon regime of black holes seem to be the ideal astrophysical laboratory to test these models against observations. In particular the continuum spectrum emitted from the accretion disk around black holes bears the imprints of the background spacetime and hence can be used as a promising probe to test the nature of strong gravity \citep{Harko:2009xf,Harko_2010,Kovacs:2010xm,Johannsen:2016uoh,Krawczynski:2018fnw,Bambi:2015kza}. 
In this work we compute the continuum spectrum from the accretion disk assuming the spacetime around the black holes to be governed by the Kerr-Sen metric. The presence of dilatonic and axionic charges modify the continuum spectrum from the Kerr scenario. The theoretical spectrum thus computed is compared with the optical data of eighty Palomar Green quasars which allows us to discern the observationally favored magnitude of the dilaton parameter and also estimate the spins of the quasars. We compute several error estimators, e.g. chi-square, Nash-Sutcliffe efficiency and index of agreement etc. to arrive at our conclusions.

The paper is organised as as follows: In \ref{S2} we describe the Einstein-Maxwell dilaton-axion (EMDA) theory and the Kerr-Sen solution. \ref{S3} is dedicated to computing the theoretical spectrum from the accretion disk in the Kerr-Sen background. The theoretical spectrum is subsequently compared with the optical observations of eighty Palomar Green quasars and the error estimators are computed in \ref{S4}. Finally we conclude with a summary of our findings with some scope for future work in \ref{S5}.

Notations and Conventions: Throughout this paper, we use (-,+,+,+) as the metric convention and will work with geometrized units taking $G=c=1$.

\section{Einstein-Maxwell dilaton-axion gravity: A brief overview }
\label{S2}
The Einstein-Maxwell dilaton-axion (EMDA) gravity \citep{Sen:1992ua,Rogatko:2002qe} is obtained when the ten dimensional heterotic string theory is compactified on a six dimensional torus $T^6$. 
The action $\mathcal{S}$ associated with EMDA gravity comprises of couplings between the metric $g_{\mu\nu}$, the $U(1)$ gauge field $A_\mu$, the dilaton field $\chi$ and the third rank anti-symmetric tensor field $\mathcal{H}_{\mu\nu\alpha}$ such that,
\begin{align}
\label{S2-1}
\mathcal{S} = \frac{1}{16\pi}\int\sqrt{-g}d^{4}x\bigg{[}~ \mathcal{R} - 2\partial_{\nu}\chi\partial^{\nu}\chi -\frac{1}{3}\mathcal{H}_{\rho\sigma\delta}\mathcal{H}^{\rho\sigma\delta} + e^{-2\chi}\mathcal{F}_{\rho\sigma}\mathcal{F}^{\rho\sigma}\bigg{]} 
\end{align}
where, $g$ is the determinant and $\mathcal{R}$ the Ricci scalar with respect to the 4-dimensional metric $g_{\mu\nu}$. In \ref{S2-1}, $\mathcal{F}_{\mu\nu}$ represents the second rank antisymmetric Maxwell field strength tensor such that $\mathcal{F}_{\mu\nu}=\nabla_\mu A_\nu-\nabla_\nu A_\mu$ while the dilaton field is denoted by $\chi$. The third rank antisymmetric tensor field $\mathcal{H}_{\rho\sigma\delta}$ in the above action can be expressed in the form,
\begin{align}
\label{S2-2}
\mathcal{H}_{\rho\sigma\delta}=\nabla_\rho B_{\sigma\delta}+\nabla_\sigma B_{\delta \rho}+\nabla_\delta B_{\rho\sigma}-(A_\rho B_{\sigma\delta}+A_\sigma B_{\delta\rho}+A_\delta B_{\rho\sigma})
\end{align}
where the second rank anti-symmetric tensor gauge field $B_{\mu\nu}$ in \ref{S2-2} is known as the Kalb-Ramond field and its cyclic permutation with $A_\mu$ represents the Chern-Simons term. 

In four dimensions $\mathcal{H}_{\mu\nu\alpha}$ is associated with the pseudo-scalar axion field $\xi$, such that,
\begin{align}
\label{S2-3}
\mathcal{H}_{\rho\sigma\delta} = \frac{1}{2}e^{4\chi}\epsilon_{\rho\sigma\delta\gamma}\partial^{\gamma}\xi
\end{align}
When expressed in terms of the axion field, \ref{S2-1} assumes the form,
\begin{align}
\label{S2-4}
\mathcal{S} &= \frac{1}{16\pi}\int\sqrt{-g}~d^{4}x\bigg{[}\mathcal{R} - 2\partial_{\nu}\chi\partial^{\nu}\chi - \frac{1}{2}e^{4\chi}\partial_{\nu}\xi\partial^{\nu}\xi + e^{-2\chi}\mathcal{F}_{\rho\sigma}\mathcal{F}^{\rho\sigma} \nonumber \\
& + \xi\mathcal{F}_{\rho\sigma}\tilde{\mathcal{F}}^{\rho\sigma}\bigg{]} 
\end{align}
By varying the action with respect to the dilaton, axion and the Maxwell fields we obtain their corresponding equations of motion. The equation of motion associated with the dilaton field is given by,
\begin{align}
\nabla_{\mu}\nabla^{\mu}\chi - \frac{1}{2}e^{4\chi}\nabla_{\mu}\xi\nabla^{\mu}\xi + \frac{1}{2}e^{-2\chi}\mathcal{F}^{2} &= 0, \label{S2-5}
\end{align}
while that of the axion is,
\begin{align}
\nabla_{\mu}\nabla^{\mu}\xi + 4\nabla_{\nu}\xi\nabla^{\nu}\xi - e^{-4\chi}\mathcal{F}_{\rho\sigma}\tilde{\mathcal{F}}^{\rho\sigma} &= 0 \label{S2-6}
\end{align} 
The Maxwell's equations coupled to the dilaton and the axion fields are given by,
\begin{align}
\nabla_{\mu}(e^{-2\chi}\mathcal{F}^{\mu\nu} + \xi\tilde{\mathcal{F}}^{\mu\nu}) &= 0,\label{S2-7}\\
\nabla_{\mu}(\tilde{\mathcal{F}}^{\mu\nu}) &= 0 \label{S2-8}
\end{align}
The solution of the axion, dilaton and the $U(1)$ gauge fields are respectively given in \cite{Ganguly:2014pwa,Sen:1992ua,Rogatko:2002qe},
\begin{align}
\xi &= \frac{q^{2}}{\mathcal{M}}\frac{a\cos\theta}{r^{2} + a^{2}\cos^{2}\theta} \label{S2-9}\\
e^{2\chi} &= \frac{r^{2} + a^{2}\cos^{2}\theta}{r(r + r_{2}) + a^{2}\cos^{2}\theta}\label{S2-10}\\
A & =\frac{qr}{\tilde{\Sigma}}\bigg(-dt +a \mathrm{sin}^2\theta d\phi\bigg)
\label{S2-11}
\end{align}
From above the non-zero components of $H_{\mu\nu\alpha}$ can also be evaluated \citep{Ganguly:2014pwa}.

The gravitational field equations are obtained when the action is varied with respect to $g_{\mu\nu}$ which yields the Einstein's equations,
\begin{align}
\mathcal{G}_{\mu\nu} = \mathcal{T}_{\mu\nu}(\mathcal{F},\chi,\xi) \label{S2-12}
\end{align}
where, $\mathcal{G}_{\mu\nu}$ is the Einstein tensor and the energy-momentum tensor $\mathcal{T}_{\mu\nu}$ on the right hand side of \ref{S2-12} is given by,
\begin{align}
\label{S2-13}
\mathcal{T}_{\mu\nu}(\mathcal{F},\chi,\xi)& = e^{2\chi}(4\mathcal{F}_{\mu\rho}\mathcal{F}_{\nu}^{\rho} - g_{\mu\nu}\mathcal{F}^{2}) - g_{\mu\nu}(2\partial_{\gamma}\chi\partial^{\gamma}\chi + \frac{1}{2}e^{4\chi}\partial_{\gamma}\xi\partial^{\gamma}\xi) 
\nonumber \\
&+ \partial_{\mu}\chi\partial_{\nu}\chi + e^{4\chi}
\partial_{\mu}\xi\partial_{\nu}\xi
\end{align}
The stationary and axisymmetric solution of the Einstein's equations corresponds to the Kerr-Sen metric \citep{Sen:1992ua} which when expressed in Boyer-Lindquist coordinates assumes the form \citep{Garcia:1995qz,Ghezelbash:2012qn,Bernard:2016wqo},
\begin{align}
\label{S2-14}
ds^{2} &= - \bigg{(}1 - \frac{2\mathcal{M}r}{\tilde{\Sigma}}\bigg{)}~dt^{2} + \frac{\tilde{\Sigma}}{\Delta}(dr^{2} + \Delta d\theta^{2}) - \frac{4a\mathcal{M}r}{\tilde{\Sigma}}\sin^{2}\theta dt d\phi  \nonumber \\
&+ \sin^{2}\theta d\phi^{2}\bigg{[}r(r+r_{2}) + a^{2} + \frac{2\mathcal{M}ra^{2}\sin^{2}\theta}{\tilde{\Sigma}}\bigg{]}
\end{align}
where,
\begin{align}
\label{S2-14a}
\tilde{\Sigma} &= r(r + r_{2}) + a^{2}\cos^{2}\theta \tag {14a}\nonumber\\
\Delta &= r(r + r_{2}) - 2\mathcal{M}r + a^{2} \tag {14b}\nonumber
\end{align}
In \ref{S2-14}, $\mathcal{M}$ is the mass, $r_{2} = \frac{q^{2}}{\mathcal{M}}e^{2\chi_{0}}$ is the dilaton parameter and $a$ is the angular momentum associated with the black hole. The dilaton parameter bears the imprints of the asymptotic value of the dilatonic field $\chi_{0}$ and the electric charge $q$ of the black hole. This charge essentially originates from the axion-photon coupling and not the in-falling charged particles since without the electric charge the field strengths corresponding to both the axion and the dilaton vanish (\ref{S2-9} and \ref{S2-10}). In such a scenario, \ref{S2-14} reduces to the Kerr metric.
We further note that in \ref{S2-14} the spin of the black hole originates from the axion field since a non-rotating black hole ($a=0$) leads to a vanishing axionic field strength (\ref{S2-9}).  
When the rotation parameter in \ref{S2-14} vanishes, the resultant spherically symmetric spacetime represents a pure dilaton black hole labelled by its mass, electric charge and the asymptotic value of the dilaton field \citep{Garfinkle:1990qj,Yazadjiev:1999xq}. 

The event horizon $r_H$ of the Kerr-Sen spacetime is obtained by solving for $\Delta=0$ such that,
\begin{align}
\label{S2-15}
r_H=\mathcal{M}-\frac{r_2}{2} +\sqrt{\bigg(\mathcal{M}-\frac{r_2}{2}\bigg)^2 - a^2}
\end{align} 
From \ref{S2-15} and the fact that $r_2$ depends on the square of the electric charge, it can be shown that $0\leq \frac{r_2}{\mathcal{M}} \leq 2$ leads to real, positive event horizons and hence black hole solutions. In this work we will be interested in this regime of $r_2$.

In the next section we will discuss accretion in the Kerr-Sen background and explicitly demonstrate the dependence of the luminosity from the accretion disk on the background spacetime.

\section{Spectrum from the accretion disk around black holes in the Kerr-Sen spacetime}
\label{S3}
The continuum spectrum emitted from the accretion disk surrounding the black holes 
is sensitive to the background metric and hence provides an interesting observational avenue to explore the signatures of the Kerr-Sen spacetime. 
In this section, we will compute the continuum spectrum from the accretion disk in a general stationary, axi-symmetric background and subsequently specialize to the Kerr-Sen metric. This in turn will enable us to probe the observable effects of EMDA gravity which can be used to distinguish it from the general relativistic scenario.

The continuum spectrum depends not only on the background metric but also on the properties of the accretion flow. In this work we will derive the continuum spectrum based on the Novikov-Thorne model which adopts the `thin-disk approximation' \citep{Novikov_Thorne_1973,Page:1974he}. In such a scenario, the accretion takes place along the equatorial plane ($\theta=\pi/2$ plane) such that the resultant accretion disk is infinitesimally thin with $\frac{h(r)}{r}\ll 1 $,  $h(r)$ being the height of the disk at a radial distance $r$. The accreting particles are assumed to maintain \emph{nearly circular} geodesics such that the azimuthal velocity $u_{\phi}$ much exceeds the radial and the vertical velocity $u_r$ and $u_z$ respectively, i.e. $u_{z} \ll u_{r} \ll u_{\phi}$. The presence of viscous stresses imparts minimal radial velocity to the accreting fluid which facilitates gradual inspiral and fall of matter
into the black hole. Since the vertical velocity is negligible, a thin accretion disk harbors \emph{no outflows}.

The energy-momentum tensor associated with the accreting fluid can be expressed as,
\begin{align}
T^{\alpha}_ {\beta}= \rho_{0}\left(1+\Pi\right)u^{\alpha}u_{\beta}+t^{\alpha}_{\beta}+u^{\alpha}q_{\beta}+q^{\alpha}u_{\beta}~,
\label{S3-1}
\end{align}
where $\rho_{0}$ is the proper density and $u^{\alpha}$ is the four velocity of the accreting particles such that the term $\rho_{0}u^{\mu}u^{\nu}$ represent the contribution to the energy-momentum tensor due to the geodesic flow. The specific internal energy of the system is denoted by $\Pi$ and the associated term denotes the contribution to the energy density due to dissipation. In \ref{S3-1}, $t^{\alpha \beta}$ and $q^{\alpha}$ respectively denote the stress-tensor and the energy flux relative to the local inertial frame and consequently $t_{\alpha \beta}u^{\alpha}=0=q_{\alpha}u^{\alpha}$. Motion of the particles along the geodesics ensures that the gravitational pull of the central black hole dominates the forces due to radial pressure gradients and hence the specific internal energy of the accreting fluid can be ignored compared to its rest energy. Therefore, the special relativistic corrections to the local thermodynamic, hydrodynamic and radiative properties of the fluid can be safely neglected compared to the general relativistic effects of the black hole \citep{Novikov_Thorne_1973,Page:1974he}. 
The loss of gravitational potential energy due to infall of matter towards the black hole  generates electromagnetic radiation which interacts efficiently with the accreting fluid before being radiated out of the system. Since the specific internal energy $\Pi\ll 1$, the accreting fluid retains no heat and the only the z-component of the energy flux vector $q^\alpha$ has a non-zero contribution to the energy-momentum tensor.

In order to compute the flux and hence the luminosity from the accretion disk we assume that the black hole undergoes steady state accretion at a rate $\dot{M}$ and the accreting fluid obeys conservation of mass, energy and angular momentum. The conservation of mass assumes the form,
\begin{align}
\label{S3-2}
\dot{M}=-2\pi \sqrt{-\tilde{g}}u^{r} \Sigma
\end{align}
where $\Sigma$ denotes the average surface density of matter flowing into the black hole and $\tilde{g}=-g_{rr}(g_{t\phi}^2-g_{tt}g_{\phi\phi})$ corresponds to the determinant of the near-equatorial metric (since we are considering motion along the $\theta=\pi/2$ plane). 
The conservation of angular momentum and energy are respectively given by,
\begin{align}
\label{S3-3}
 \frac{\partial}{\partial r}\left[\dot{M}\mathcal{L}-2\pi \sqrt{-\tilde{g}}W^r_\phi \right]=4\pi \sqrt{-\tilde{g}} F \mathcal{L}~,~~~\rm{and } 
\end{align}
\begin{align}
\label{S3-4}
\frac{\partial}{\partial r}\left[\dot{M}\mathcal{E}-2\pi \sqrt{-\tilde{g}}\Omega W^r_\phi \right]=4\pi \sqrt{-\tilde{g}} F \mathcal{E} 
\end{align}
where $\Omega$, $\mathcal{E}$ and $\mathcal{L}$ are the angular velocity, the specific energy and the specific angular momentum of the accreting fluid. In a stationary and axi-symmetric spacetime $\mathcal{E}$ and $\mathcal{L}$ are conserved and can be expressed in terms of the metric coefficients such that,
\begin{align}
\label{S3-5a}
\mathcal{E}&=\frac{-g_{tt}-\Omega g_{t\phi}}{\sqrt{-g_{tt}-2\Omega g_{t\phi}-\Omega^2 g_{\phi\phi}}}~,
\end{align} 
\begin{align}
\label{S3-5b}
\mathcal{L}&=\frac{\Omega g_{\phi\phi}+g_{t\phi}}{\sqrt{-g_{tt}-2\Omega g_{t\phi}-\Omega^2 g_{\phi\phi}}}~.
\end{align} 
and the angular velocity $\Omega$ is given by,
\begin{align}
\label{S3-6}
\Omega=\frac{d\phi}{dt}=\frac{-g_{t\phi,r}\pm \sqrt{\left\lbrace-g_{t\phi,r}\right\rbrace^2-\left\lbrace g_{\phi\phi,r}\right\rbrace \left\lbrace g_{tt,r}\right\rbrace}}{g_{\phi\phi,r}}
\end{align}
Since the motion is along the equatorial plane $\mathcal{E}$ and $\mathcal{L}$ are only functions of the radial coordinate and the $g_{\theta \theta}$ component does not contribute to the conserved quantities. 

In \ref{S3-3} and \ref{S3-4}, $F$ denotes the flux of radiation generated by the accretion process and is given by,
\begin{align}
\label{S3-6}
F \equiv \left\langle q^z(r,h)\right\rangle=\left\langle -q^z(r,-h)\right\rangle 
\end{align}
while $W^{r}_{\phi}$ is associated with the time and height averaged stress tensor in the local rest frame of the accreting particles, i.e., 
\begin{align}
\label{S3-7}
W^\alpha_\beta=\int^h_{-h}dz\left\langle t^\alpha_\beta\right\rangle
\end{align}

By manipulating the conservation laws an analytical expression for the flux $F$ from the accretion disk can be obtained, such that \citep{Page:1974he},
\begin{align}
\label{S3-8}
F = \frac{\dot{M}}{4\pi\sqrt{-\tilde{g}}}f ~~~\rm{where}
\end{align}
\begin{align}
\label{S3-9}
f=-\frac{\Omega_{,r}}{(\mathcal{E}-\Omega \mathcal{L})^2}\left[\mathcal{EL}-\mathcal{E}_{ms}\mathcal{L}_{ms}-2\int_{r_{ms}}^r \mathcal{LE}_{,r^\prime}dr^\prime \right]
\end{align}
While deriving \ref{S3-8} the viscous stress $W^r_\phi$ is assumed to vanish at the marginally stable circular orbit $r_{ms}$, such that after crossing this radius the azimuthal velocity of the accreting fluid vanishes and radial accretion takes over. The last stable circular orbit $r_{ms}$, corresponds to the inflection point of the effective potential in which the accreting particles move. The effective potential $V_{eff}$ is given by \citep{Banerjee:2019sae},
\begin{align}
\label{S3-10}
V_{\rm eff}(r)=\frac{\mathcal{E}^2g_{\phi\phi}+2\mathcal{EL}g_{t\phi}+\mathcal{L}^2g_{tt}}{g_{t\phi}^2-g_{tt}g_{\phi\phi}}-1
\end{align}
and $r_{ms}$ is obtained by solving for $V_{\rm eff}=\partial _{r}V_{\rm eff}=\partial _{r}^{2}V_{\rm eff}=0$. It is however important to check that the orbits are vertically stable. This has been discussed by \cite{Ono:2016lql} and following their stability criterion we have verified that the vertical stability of the orbits hold true for the Kerr-Sen metric.
In \ref{S3-9} $\mathcal{E}_{ms}$ and $\mathcal{L}_{ms}$ denote the energy and angular momentum at the marginally stable circular orbit. 

Since the electromagnetic radiation emitted due to loss of gravitational potential energy  undergoes repeated collisions with the accreting fluid, a thermal equilibrium between  matter and radiation is established. This renders the accretion disk to be geometrically thin and optically thick such that it emits locally as a black body. Therefore, the temperature profile is given by the Stefan-Boltzmann law, i.e., $T(r)=\left(\mathcal{F}(r)/\sigma\right)^{1/4}$ ($\sigma$ being the Stefan-Boltzmann constant) and $\mathcal{F}(r)=F(r)c^6/G^2\mathcal{M}^2$ (where $F(r)$ is given by \ref{S3-8} and \ref{S3-9}). Hence the accretion disk emits a Planck spectrum at every radius with temperature $T(r)$.
Finally, the luminosity $L_{\nu}$ emitted by the disk at an observed frequency $\nu$ is obtained by integrating the Planck function $B_\nu(T(r))$ over the disk surface,
 \begin{align}
\label{S3-11}
L_{\nu}=8\pi^2 r_{\rm g}^2\cos i  \int_{x_{\rm ms}}^{x_{\rm out}}\sqrt{g_{rr}} B_{\nu}(T)x dx;
\qquad
B_\nu (T)&=\frac{2h\nu^3/c^2}{{\rm exp}\left(\frac{h\nu}{z_{\rm g} kT}\right)-1}
\end{align}
where, $x = r/r_{\rm g}$ is the radial coordinate expressed in units of the gravitational radius $r_{\rm g} = G\mathcal{M}/c^{2}$ and $``i"$ is the inclination angle between the line of sight and the normal to the disk. In \ref{S3-11} $z_{\rm g}$ is the gravitational redshift factor given by,
\begin{align}
\label{S3-12}
z_{\rm g}=\mathcal{\tilde{E}}\frac{\sqrt{-g_{tt}-2\Omega g_{t\phi}-\Omega^2 g_{\phi\phi}}}{\mathcal{\tilde{E}}-\Omega \mathcal{\tilde{L}}} 
\end{align} 
which is associated with the change in the frequency suffered by the photon while travelling from the emitting material to the observer \citep{Ayzenberg:2017ufk}. In \ref{S3-12}, $\mathcal{\tilde{E}}$ and $\mathcal{\tilde{L}}$ are specific energy and specific angular momentum associated with the outgoing photon.

\begin{figure*}
\hspace{-2.1cm}
\includegraphics[scale=0.69]{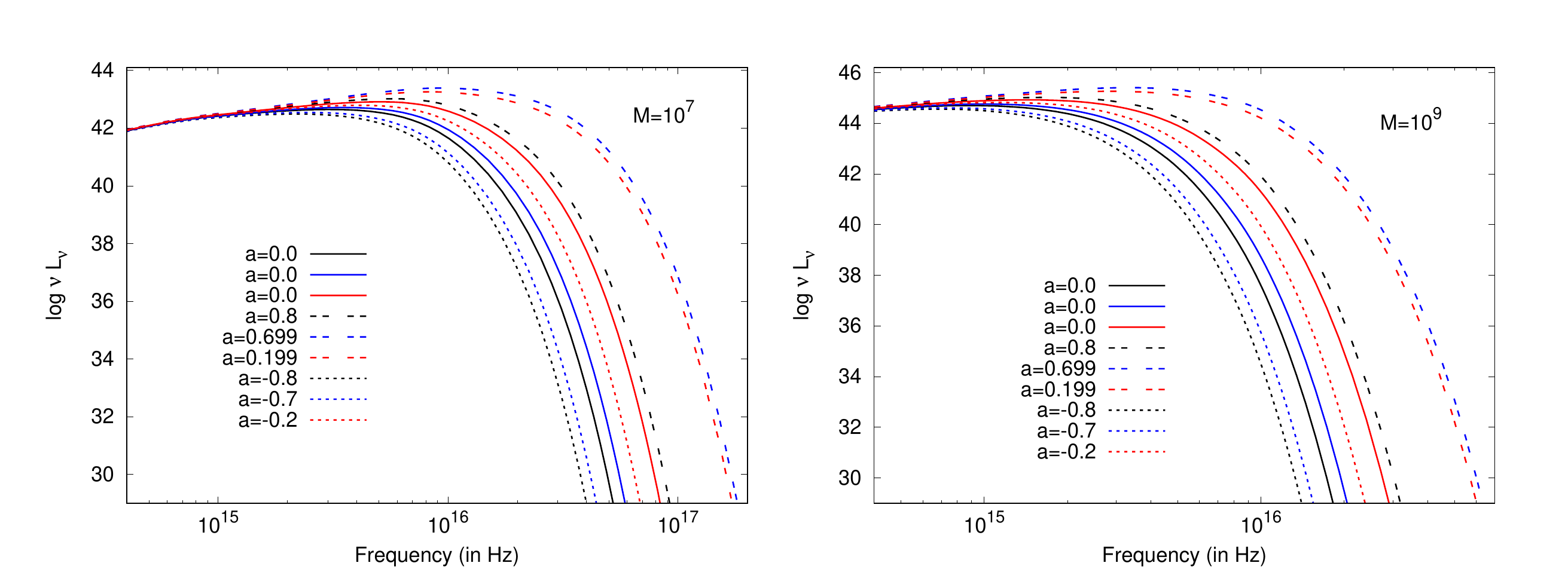}
\caption{The above figure depicts the variation of the theoretically derived luminosity from the accretion disk with frequency for two different masses of the supermassive black holes, namely, $M=10^7 M_\odot$ and $M=10^9 M_\odot$. 
For both the masses, the black lines represent $r_2=0$, while the blue and red lines correspond to $r_2=0.6$ and $r_2=1.6$ respectively. For a given $r_2$, prograde spins are denoted by dashed lines, non-rotating black holes are denoted by solid lines while their retrograde counterparts are illustrated by the dotted lines. The Schwarzschild scenario is depicted by the solid black line. The accretion rate is assumed to be $0.1 \dot{M}_{Edd}$ and the inclination angle is taken to be $\cos i=0.8$. For more discussions see text.
}
\label{F1}
\end{figure*}

We have thus arrived at an analytical expression for the luminosity from the accretion disk given by \ref{S3-11}. We note that it depends on the background metric through the energy, angular momentum, angular velocity and the radius of the marginally stable circular orbit, where in the present work we will consider the metric components corresponding to the Kerr-Sen spacetime given by \ref{S2-14}. Apart from the metric parameters, the spectrum is also sensitive to the mass of the black hole, the accretion rate and the inclination angle of the disk to the line of sight. 

\ref{F1} depicts the variation of the theoretically derived luminosity from the accretion disk with the frequency, for two different masses of the supermassive black holes, viz, $\mathcal{M}=10^9 M_\odot$ and $\mathcal{M}=10^7 M_\odot$. The accretion rate is assumed to be $ 0.1\dot{M}_{Edd}$ (where $ 1 \dot{M}_{Edd}=1.4\times 10^{17}
\mathcal{M}/M_\odot ~\rm gm ~s^{-1}$) while the inclination angle is considered to be $\cos i=0.8$. We note that the spectrum from a lower mass black hole peaks at a higher frequency which can be ascribed to the $T \propto \mathcal{M}^{-1/4}$ dependence of the temperature $T$ for a multi-color black body spectrum with the black hole mass \citep{Frank:2002}. It is evident from \ref{F1} that the metric parameters $r_2$ and $a$ substantially affect the luminosity from the accretion disk, specially at high frequencies. We recall from previous discussion that for the existence of the event horizon, the spin of the black hole should lie in the range: $-(1-\frac{r_2}{2})\leq a \leq (1-\frac{r_2}{2})$ for a given $r_2$ (\ref{S2-15}), where $a$ and $r_2$ are expressed in units of $\mathcal{M}$, which will be considered throughout the remaining discussion. We study three different dilaton parameters in \ref{F1}, $r_2=0$ (black lines), $r_2=0.6$ (blue lines) and $r_2=1.6$ (red lines). For each dilaton parameter we consider non-rotating black holes (solid lines) as well as the prograde (dashed lines) and retrograde (dotted lines) spins in the allowed range. We note that for a given $r_2$ the disk luminosity associated with prograde black holes is maximum followed by the luminosity corresponding to their non-spinning and the retrograde counterparts. Similarly, for a given spin, accretion disks around dilaton black holes are more luminous  compared to the general relativistic scenario.

\section{Numerical Analysis}
\label{S4}
In this section we compute the optical luminosity of a sample of Palomar Green (PG) quasars \citep{Schmidt:1983hr,Davis:2010uq} assuming the thin accretion disk model discussed in the last section. The optical luminosity $L_{opt}\equiv \nu L_\nu$ is evaluated at the wavelength 4861\AA\ \citep{Davis:2010uq} and compared with the corresponding observed values \citep{Davis:2010uq} which in turn allows us to discern the observationally favored magnitude of the dilaton parameter and the black hole spins. These quasars have independent mass measurements based on the method of reverberation mapping \citep{Davis:2010uq,Kaspi:1999pz,Kaspi:2005wx,Boroson:1992cf,Peterson:2004nu} while the accretion rates of the quasars are reported in \cite{Davis:2010uq}. 
In \citet{Davis:2010uq}, the accretion rates of the quasars are estimated assuming   a stellar-atmosphere-like model for the disk (the so called TLUSTY model) and by considering the black hole spin to be $a=0.9$ (which is their base model). However, they have mentioned the degree of variation that can be expected in the accretion rates if a different disk model is used (e.g. the local black body model) or a different black hole spin is considered. It turns out that if the TLUSTY model is used with $a=0$, the quasars with higher mass exhibit a 40\% increase in accretion rate while the ones with lower mass assumes an accretion rate 10\% higher (compared to the ones estimated based on TLUSTY model with $a=0.9$, i.e., the base model). On the other hand, if black body models are considered with $a=0.9$, then the accretion rates turn out to be lower by 10\% to 20\% (compared to the ones reported in \citet{Davis:2010uq}) for all quasars. Likewise, if blackbody model with $a=0$ is considered, then the high mass quasars turn out to have an increase in the accretion rate by 40\% while the low mass quasars exhibit a decline in the accretion rate by 20\% (compared to the base model). This discussion reveals that whatever be the choice of the disk model or the black hole spin, the accretion rate of the PG quasars can at most vary between 80\% to 140\% of the rate reported in \citet{Davis:2010uq}. This factor will be taken into account while estimating the most favored magnitude of $r_2$ from the quasar optical data.

The quasar sample considered in \citet{Davis:2010uq} are nearly face-on systems and hence the inclination angle $i$ is expected to lie in the range: $\cos i \in \left(0.5,1\right)$ \citep{Antonucci:1993sg,Davis:2010uq,Wu:2013zqa}. This is in agreement with the results of \cite{2017Ap&SS.362..231P} who estimated the inclination angle of some of these quasars using the degree of polarisation of the scattered radiation from the accretion disk. 
Although a uniform inclination angle of $\cos i\sim 0.8$ is assumed in \citet{Davis:2010uq}, in this work we allow the inclination angle to vary in the aforesaid range to understand the impact of $\cos i$ on our estimate of $r_2$.

\begin{figure*}
\hspace{-2.1cm}
\includegraphics[scale=0.63]{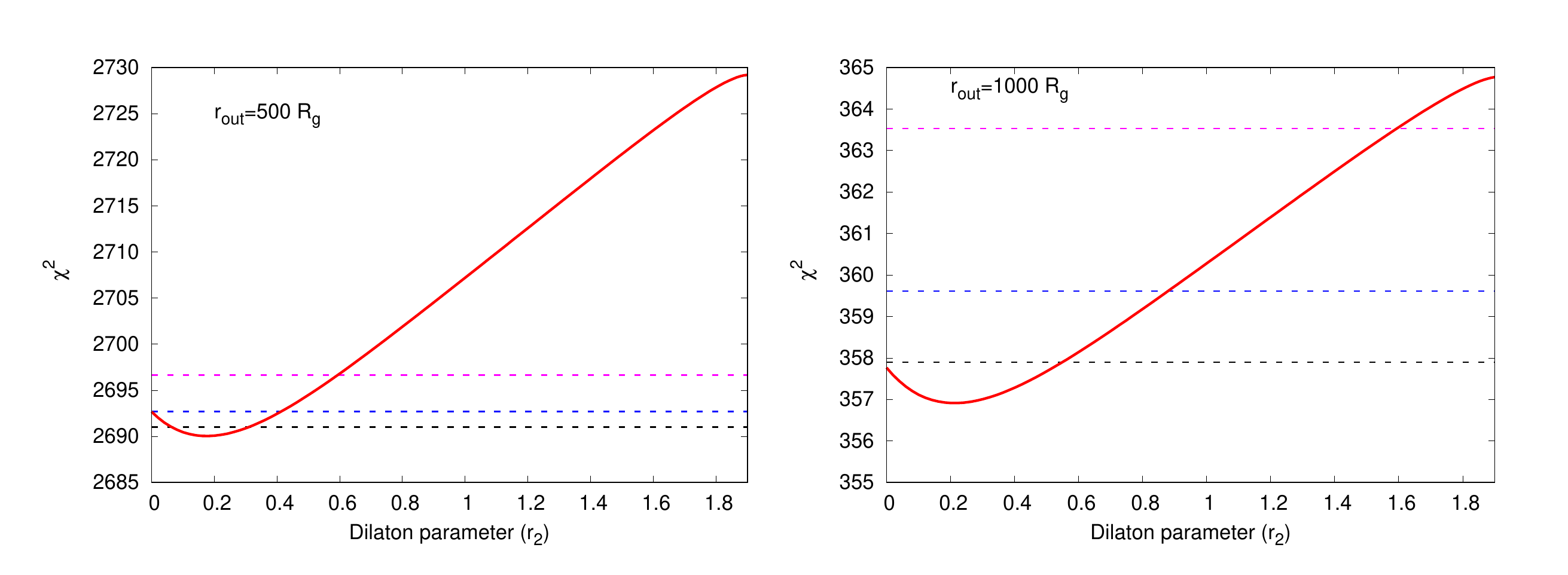}
\caption{ The above figure depicts the variation of $\chi^2$ with the dilaton parameter $r_2$ for the sample of eighty PG quasars. In the left panel, the theoretical luminosity and hence the $\chi^2$ is computed with $r_{out}=500r_g$ while $r_{out}=1000r_g$ is used to compute $\chi^2$ in the right panel. Interestingly, the $\chi^2$ minimizes for the same dilaton parameter ($r_2\sim 0.2$) irrespective of the choice of $r_{out}$, although the values of $\chi^2$ vary in different range if a different choice of $r_{out}$ is considered. The $\Delta \chi^2$ corresponding to 68\%, 90\% and 99\% confidence intervals (for a single parameter) are also plotted in the figure with black, blue and magenta dashed lines, respectively. For discussions see text.}

\label{Fig_2}
\end{figure*}

The bolometric luminosities of these quasars have been estimated \citep{Davis:2010uq} based on the observed data in the optical \citep{1987ApJS...63..615N}, UV \citep{Baskin:2004wn}, far-UV \citep{Scott:2004sv}, and soft X-rays \citep{Brandt:1999cm}. The error in bolometric luminosity receives dominant contribution from the far-UV extrapolation since the uncertainty in the UV luminosity supercedes other sources of error (e.g., optical or X-ray variability) \citep{Davis:2010uq}. Moreover, the UV part of the spectral energy distribution (SED) is contaminated by components other than the accretion disk since physical mechanisms e.g. advection, a Comptonizing coronae, etc. may redistribute the UV flux to the X-ray frequencies \citep{Davis:2010uq}. 
Therefore, although the maximum emission from the accretion disk for quasars generally peaks in the optical/UV part of the spectrum, disentangling the role of the metric from UV observations becomes difficult due to the aforesaid reasons. This motivates us to dwell in the optical domain and compare the optical observations of quasars with the corresponding theoretical estimates.

The requirement for the quasars to possess a real, positive event horizon imposes the constraint $0\leq r_2\leq 2$ on the dilaton parameter. Also, for a given $r_2$, the spin $a$ of the black hole can assume values between, $-(1-\frac{r_2}{2})\leq a \leq (1-\frac{r_2}{2})$. In order to arrive at the most favored dilaton parameter we proceed in the following way:
\begin{itemize}
\item We consider a quasar in the sample with known $\mathcal{M}$ (from reverberation mapping) and a given $r_2$. We vary its $\dot{M}$ between $0.8-1.4$ times the accretion rate reported in \citet{Davis:2010uq}. For a given $\dot{M}$, we vary the inclination angle $i$, such that $\cos i $ ranges from $0.5-1$ and finally for a chosen $\dot{M}$ and $\cos i$, the spin is varied between $-(1-\frac{r_2}{2})\leq a \leq (1-\frac{r_2}{2})$. For every combination of $\dot{M}$, $\cos i$ and $a$ the theoretical optical luminosity is computed at 4861\AA. The values of $\dot{M}$, $\cos i$ and $a$ that best reproduces the observed optical luminosity, is considered to be the most favored magnitude of accretion rate, inclination angle and spin for the chosen quasar at the given $r_2$.    

\item We repeat the above procedure for all the quasars in the sample for the aforesaid $r_2$. This assigns a specific spin, accretion rate and $\cos i$ to each of the quasars, for the given $r_2$. 

\item We now vary $r_2$ in the theoretically allowed range,  ($0\leq r_2\leq 2$) and repeat the above two procedures. This ensures that for every $r_2$, the sample of quasars are associated with $\dot{M}$, $\cos i$ and $a$ that minimizes the error between the theoretical and the observed optical luminosities.

\end{itemize}
It may be worthwhile to compare our approach with that of \citet{Tripathi:2020qco} where the authors test the Kerr nature of the black hole LMC X-1 using the Continuum-Fitting Method. They explore the prospect of the Johannsen metric \citep{Johannsen:2015pca} in describing the Continuum spectrum of LMC X-1 compared to the Kerr metric in general relativity. In order to arrive at their results they fit seventeen spectra of the source (suitable for applying the Continuum Fitting method) obtained from the RXTE PCA with the model TBABS $\times$ (SIMPL $\times$ NKBB) and conclude that constraining $a$ and $\alpha_{13}$ (the deformation parameter corresponding to the Johannsen metric) simultaneously from the observed continuum spectrum is quite non-trivial. They report strong correlation between the spin and the deformation parameter and as a consequence their constraints are weak. They further comment that it is easier to break the degeneracy between the metric parameters if one uses the reflection spectrum instead of the thermal continuum since the former has several characteristic features compared to the simple shape of the latter. A combination of thermal and reflection spectrum is expected to impose more stringent constraints on the background spacetime.

Our approach is different from \cite{Tripathi:2020qco} since we consider optical luminosities of multiple sources to constrain the background metric and not multiple spectra of the same source. This is because, quasars are multi-component systems comprising of the accretion disk, the corona, the jet and the dusty torus, each contributing to the spectral energy distribution (SED), which makes discerning the effect of the background metric from the SED extremely challenging. To extract information regarding the background spacetime one needs to consider only emissions from regions very close to the black hole which motivates us to model the emissions from the accretion disk. Since the Novikov-Thorne model is quite successful in describing the emission from the accretion disk despite its simplistic assumptions, and constructing a purely theoretical model of the entire SED is extremely challenging, we attempt to understand the role of the Kerr-Sen metric in explaining the optical data of the quasars through this prefatory approach, as a first step.

We further mention that while deriving the most favored magnitude of the dilaton parameter $r_2$ as described earlier, we allow each quasar to have their own spin, accretion rate and inclination angle but consider the same $r_2$ for all quasars. In principle the quasars can have their own electric charges (although the electric charge of dilaton-axion black holes originate from the axion-photon coupling and not the in-falling charged particles) and if we were fitting multiple spectra of the same source then we could have kept all the four parameters ($\dot{M}$, $\cos i$, $a$ and $r_2$) free and obtain their best fit values and errors within various confidence intervals for each source, as done in \cite{Tripathi:2020qco}. Here instead, we follow the approach adopted by \cite{Pei:2016kka}, where the authors constrain the deformation parameters of the Johannsen metric (which are also expected to vary amongst different sources) by comparing the Blandfor-Znajeck jet power and radiative efficiencies (which are sensitive to the background metric) of six different microquasars with the corresponding observed jet power and black hole spin.

The common dilaton parameter of the quasars considered here in a way denote the average of the dialton parameters for all the eighty quasars. Therefore, if a non-zero dilaton parameter is favored from the observational data, it implies that on an average the chosen sample of quasars are more likely axion-dilaton black holes instead of Kerr black holes.

\subsection{Error estimators}\label{S4-1}
In this section we discuss various error estimators which enable us to find the dilaton parameter $r_{2}$ minimizing the error between the theoretical and the observed optical luminosities.

\begin{itemize}
\item {\textbf{Chi-square $\boldsymbol {\chi^{2}}~$:~}
We consider the sample of quasars with observed optical luminosities $\{\mathcal{O}_{k}\}$ and errors $\{\sigma_{k}\}$. The theoretical estimates of the optical luminosity corresponds to $\left\lbrace \mathcal{T}_{k}\left(r_{2},\left\lbrace a_{(j)}, \cos i_{(j)},\dot{M}_{(j)}\right\rbrace\right)\right\rbrace$ for a given $r_2$ (where $\left\lbrace a_{(j)}, \cos i_{(j)}, \dot{M}_{(j)}\right\rbrace$ denotes the set of spin, inclination angle and accretion rate associated with the $k^{th}$ quasar that minimizes the error between the theoretical and the observed optical luminosity for the chosen $r_2$, as discussed in the last section). With this, the chi-square ($\chi ^{2}$) of the distribution can be defined as,

\begin{align}
\label{chi2}
\chi ^{2}\bigg(r_{2},\left \lbrace a_{(j)}, \cos i_{(j)},\dot{M}_{(j)}\right\rbrace \bigg)= \sum _{k}\frac{\left\{\mathcal{O}_{k}-\mathcal{T}_{k} \bigg(r_{2},\left \lbrace a_{(j)}, \cos i_{(j)},\dot{M}_{(j)}\right\rbrace \bigg) \right\}^{2}}{\sigma _{k}^{2}}.
\end{align}
The errors associated with the optical luminosities are not explicitly reported in \citet{Davis:2010uq}. However, we have already mentioned that the error in optical luminosity can be ignored compared to the error in bolometric luminosity which receives maximum contribution from the far-UV extrapolation since the uncertainty in the UV luminosity dominates over other sources of error (e.g., optical or X-ray variability \citep{Davis:2010uq}). Therefore, we consider the errors in the bolometric luminosity reported in \citet{Davis:2010uq}, as the maximum possible error in the estimation of the optical luminosity.

From the definition of $\chi^2$ in \ref{chi2}, it is clear that the magnitude of $r_2$ which minimizes $\chi^2$ is most favored by the observations. 
It is however important to note that we do not minimize $\chi^2$ for all the parameters in the theoretical luminosity (e.g. $r_2$, $a$, $\dot{M}$ and $\cos i$) simultaneously. This is because we are interested in understanding the most favored magnitude of $r_2$ from the quasar optical data. In such a scenario, one divides the total number of model parameters into two classes: the ``interesting parameters" whose values one need to estimate simultaneously from $\chi^2$ minimization and the ``uninteresting parameters"  whose values are obtained by minimizing the $\chi^2$ function by keeping the ``interesting parameters" fixed to different values \citep{1976ApJ...210..642A}. The confidence intervals or the increment in $\chi^2$ ($\Delta \chi^2$) from the minimum value ($\chi^2_{min}$) are determined based on the number of ``interesting parameters" which in our case in one, i.e., $r_2$. Therefore, $\Delta \chi^2$ for one ``interesting parameter" corresponding to 68\%, 90\% and 99\% confidence intervals are 1, 2.71 and 6.63 \citep{1976ApJ...210..642A}.

\begin{figure*}
\hspace{-2.1cm}
\includegraphics[scale=0.63]{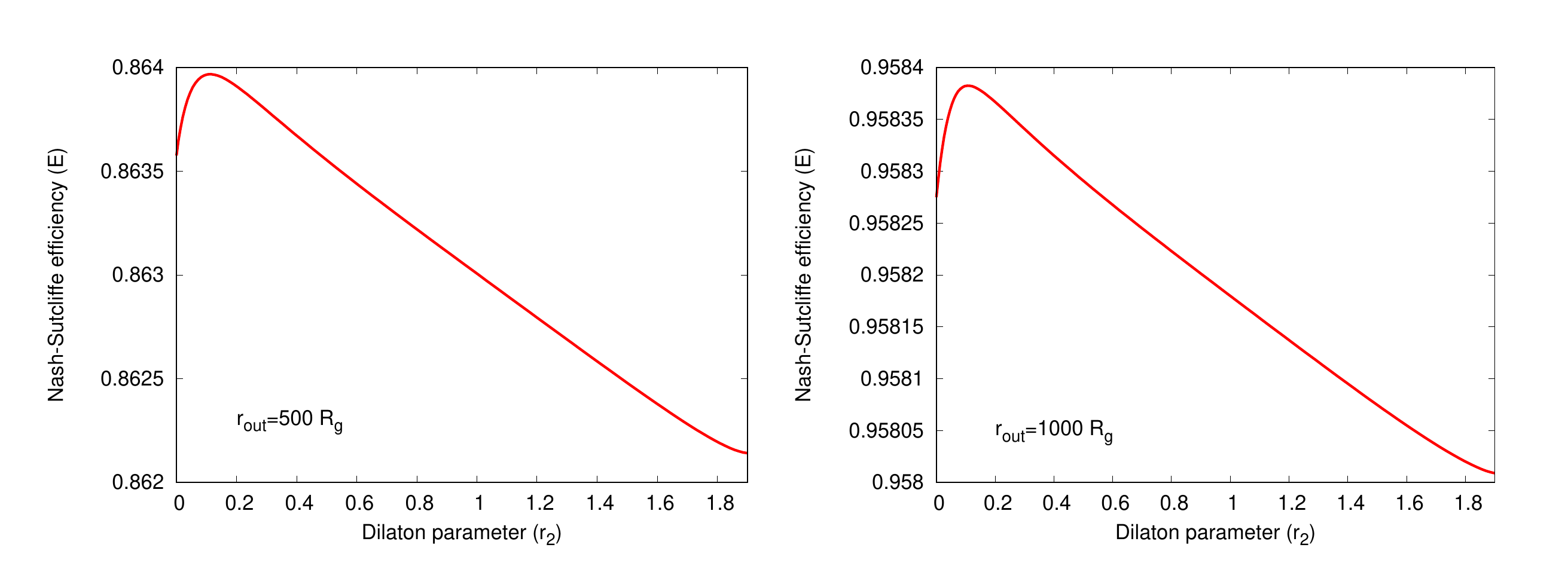}
\caption{The above figure depicts the variation of the Nash-Sutcliffe Efficiency E with the dilaton parameter $r_2$. The Nash-Sutcliffe Efficiency computed with $r_{out}=500r_g$ is shown in the left panel while $E$ computed with $r_{out}=1000r_g$ is shown in the right panel. We note that irrespective of the choice of $r_{out}$, $E$ maximizes for $r_2\sim 0.1$. For more discussion, see text.}
\label{Fig_3}
\end{figure*}

In \ref{Fig_2} we plot the variation of $\chi^{2}$ with the dilaton parameter $r_{2}$. Since the theoretical luminosity depends not only on the inner radius of the disk $r_{ms}$ but also on the outer radius $r_{out}$ (\ref{S3-11}), we consider two possible choices of $r_{out}$, namely $r_{out}=500r_g$ and $r_{out}=1000r_g$ in computing the theoretical luminosity and subsequently the $\chi^2$. These correspond to the left and the right panels of \ref{Fig_2} respectively. 
The figure clearly reveals that $r_2\sim 0.2$ minimizes the $\chi^2$ irrespective of the choice of $r_{out}$, signalling that axion-dilaton black holes with mild dilaton charges are more favored by quasar optical data. The confidence limits associated with 68\%, 90\% and 99\% intervals denoted respectively by the black, blue and magenta dashed lines, are also depicted in \ref{Fig_2}. 
We note that if we consider $r_{out}=500r_g$ then $0.05\lesssim r_2 \lesssim 0.3$ lies within 1-$\sigma$ confidence level, while extending $r_{out}$ to $1000r_g$ brings $0 \lesssim r_2 \lesssim 0.5$ within the 1-$\sigma$ interval.

It is important to recall that there are restrictions in the magnitude of $r_2$ and $a$ (as discussed in the last section) and one cannot assume arbitrary values of these parameters. Consequently we cannot use reduced chi-square $\chi ^{2}_{Red}$, where $\chi ^{2}_{Red}=\chi ^{2}/\nu$, ($\nu$ being the degrees of freedom) as an error estimator, since the definition of the degrees of freedom becomes ambiguous in such cases \citep{Andrae:2010gh}. Hence we have analyzed the error between the theoretical and the observed optical luminosities using $\chi^2$ as the error estimator.

In what follows we consider a few more error estimators in order to verify our conclusion.  
}


\item{\textbf{Nash-Sutcliffe Efficiency:~} This particular error estimator denoted by $E$  \citep{NASH1970282,WRCR:WRCR8013,2005AdG.....5...89K} is associated with the the sum of the squared differences between the observed and the theoretical values normalized by the variance of the observed values. 
The functional form of the Nash-Sutcliffe Efficiency is given by,
\begin{align}
\label{NSE}
E\bigg(r_{2},\left \lbrace a_{(j)}, \cos i_{(j)},\dot{M}_{(j)}\right\rbrace \bigg)=1-\frac{\sum_{k}\left\{\mathcal{O}_{k}-\mathcal{T}_{k}\bigg(r_{2},\left \lbrace a_{(j)}, \cos i_{(j)},\dot{M}_{(j)}\right\rbrace \bigg)\right\}^{2}}{\sum _{k}\left\{\mathcal{O}_{k}-\mathcal{O}_{\rm av}\right\}^{2}}
\end{align}
where, $\mathcal{O}_{\rm av}$ represents the mean value of the observed optical luminosities of the PG quasars.

In contrast to $\chi^{2}$, the dialton parameter which maximizes $E$ is most favored by observations.
Interestingly $E$ can range from $-\infty ~\rm to ~ 1$. Negative $E$ indicates that the average of the observed data explains the observation better than the theoretical model. Similarly, $E \sim 1$ represents the ideal model which predicts the observations with great accuracy \citep{Goyal}. The variation of the Nash-Sutcliffe efficiency with $r_2$ is illustrated in \ref{Fig_3} for both choices of $r_{out}$. We note that $E$ maximizes when $r_2\sim 0.1$ (close to $\chi^2_{min}$ in \ref{Fig_2}) thereby supporting our earlier conclusion derived from chi-square analysis. The maximum values of Nash-Sutcliffe efficiency are $E_{max}\sim 0.864$ and $E_{max}\sim 0.9584$ for $r_{out}=500r_g$ and $r_{out}=1000r_g$ respectively, which show that this is a satisfactory model representing the data \citep{Goyal}  
}.


\item{\textbf{Modified Nash-Sutcliffe Efficiency $\bf E_{1}$ :~}
In order to overcome the oversensitivity of the Nash-Sutcliffe efficiency to higher values of the optical luminosity, (which arises due to the presence of the square of the error in the numerator) a modified version of the same is proposed \citep{WRCR:WRCR8013}which is given by, 
\begin{align}\label{E1}
E_{1}\bigg(r_{2},\left \lbrace a_{(j)}, \cos i_{(j)},\dot{M}_{(j)}\right\rbrace \bigg)&=1-\frac{\sum_{k}\big|\mathcal{O}_{k}-\mathcal{T}_{k}\bigg(r_{2},\left \lbrace a_{(j)}, \cos i_{(j)},\dot{M}_{(j)}\right\rbrace \bigg)\big|}{\sum _{k}\big|\mathcal{O}_{k}-\mathcal{O}_{\rm av}\big|}
\end{align}

\begin{figure*}
\hspace{-2.1cm}
\includegraphics[scale=0.63]{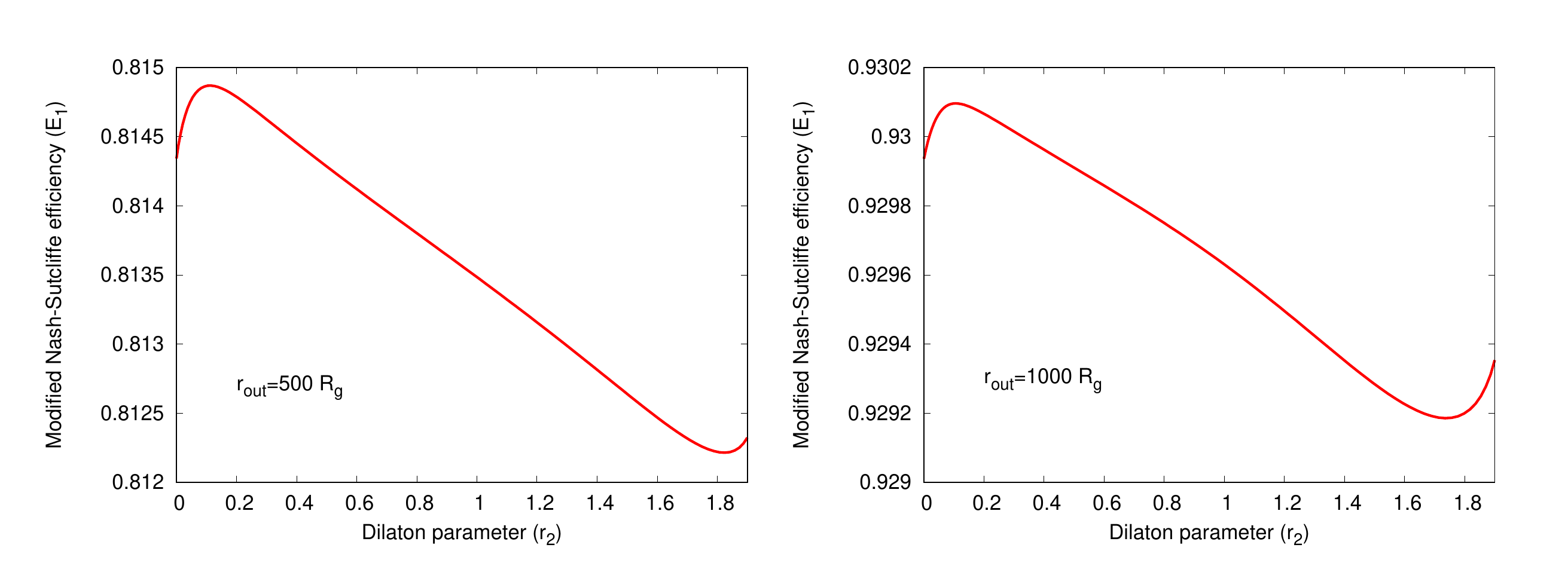}
\caption{The above figure shows the variation of the modified Nash-Sutcliffe Efficiency $E_1$ with the dilaton parameter $r_2$. As before, $E_1$ computed with $r_{out}=500r_g$ is shown in the left panel while $E_1$ computed with $r_{out}=1000r_g$ is illustrated in the right panel. As reported in \ref{Fig_3}, irrespective of the choice of $r_{out}$, $E_1$ maximizes for $r_2\sim 0.1$. For more discussion, see text.
}
\label{Fig_4}
\end{figure*}

This modified estimator exhibits an enhanced sensitivity towards the lower values of the observed optical luminosities. 
Similar to Nash-Sutcliffe Efficiency, the most favorable $r_{2}$ maximizes its modified version as well. It is observed from the \ref{Fig_4} that this maximization occurs at $r_{2}\sim 0.1$ irrespective of the choice of $r_{out}$ which in turn supports our earlier findings.
}


\item{\textbf{Index of Agreement and its modified form :~}
The Nash-Sutcliffe efficiency and its modified form turns out to be insensitive towards
the differences between the theoretical and the observed luminosities from the corresponding observed mean \citep{WRCR:WRCR8013}. This is overcome by introducing the index of agreement $d$ \citep{willmott1984evaluation, doi:10.1080/02723646.1981.10642213,2005AdG.....5...89K} where,
\begin{align}\label{d}
&d\bigg(r_{2},\left \lbrace a_{(j)}, \cos i_{(j)},\dot{M}_{(j)}\right\rbrace \bigg)=
\nonumber \\
&1-\frac{\sum_{k}\left\{\mathcal{O}_{k}-\mathcal{T}_{k}\bigg(r_{2},\left \lbrace a_{(j)}, \cos i_{(j)},\dot{M}_{(j)}\right\rbrace \bigg)\right\}^{2}}{\sum _{k}\left\{ \big|\mathcal{O}_{k}-\mathcal{O}_{\rm av}\big|+\big|\mathcal{T}_{k}\bigg(r_{2},\left \lbrace a_{(j)}, \cos i_{(j)},\dot{M}_{(j)}\right\rbrace \bigg)-\mathcal{O}_{\rm av}\big|\right\}^{2}}
\end{align}
and $\mathcal{O}_{\rm av}$ refers to the average value of the observed luminosities. The denominator, known as the potential error, is related to the maximum value by which each pair of observed and predicted luminosities differ from the average observed luminosity.

\begin{figure*}
\hspace{-2.1cm}
\includegraphics[scale=0.63]{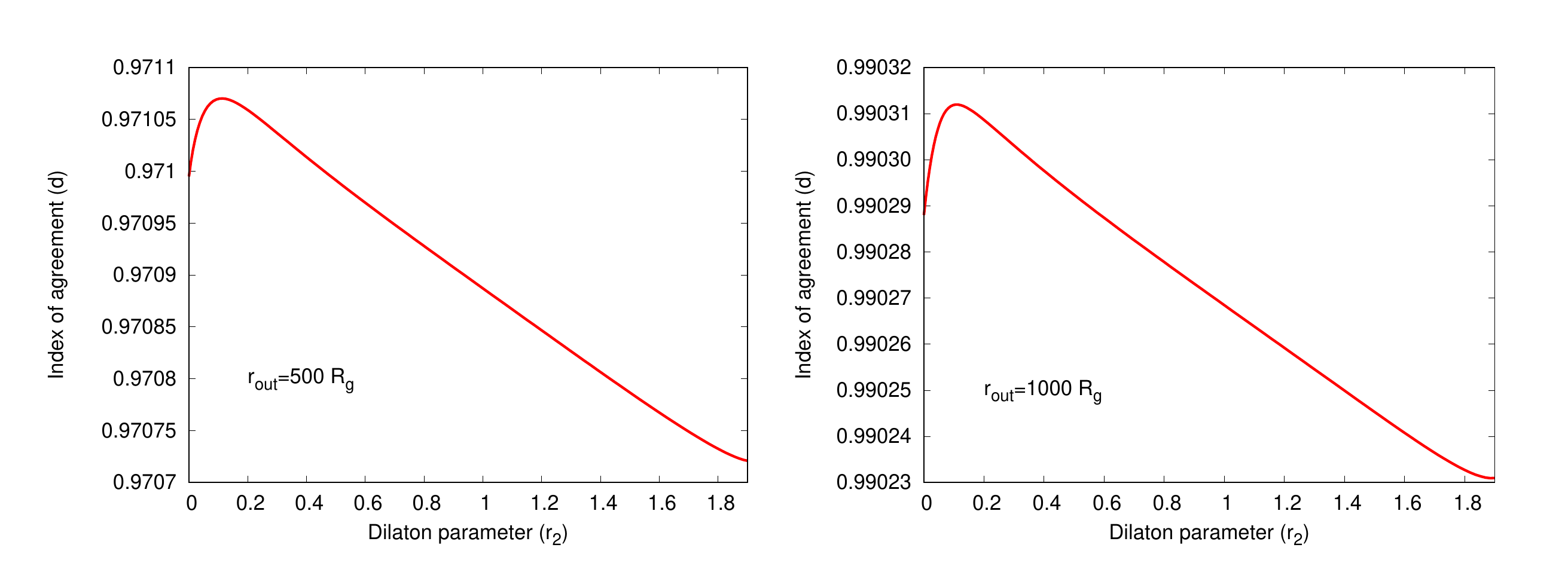}
\caption{The above figure shows the variation of the index agreement $d$ with the dilaton parameter $r_2$. The left panel depicts the index of agreement computed with $r_{out}=500r_g$, while the right panel shows the index of agreement computed with $r_{out}=1000r_g$. Consistent with the results obtained for the Nash-Sutcliffe efficiency and its modified form, the index of agreement also maximizes for $r_2\sim 0.1$, irrespective of the choice of $r_{out}$. See text for more discussions.
}
\label{Fig_5}
\end{figure*}

The presence of the squared luminosities in the numerator makes the index of agreement oversensitive to higher values of the optical luminosity.
Therefore, a modified version of the same denoted by $d_1$ is proposed which assumes the form, 
\begin{align}\label{d1}
&d_{1}\bigg(r_{2},\left \lbrace a_{(j)}, \cos i_{(j)},\dot{M}_{(j)}\right\rbrace \bigg)=\nonumber \\
&1-\frac{\sum_{k}\big|\mathcal{O}_{k}-\mathcal{T}_{k}\bigg(r_{2},\left \lbrace a_{(j)}, \cos i_{(j)},\dot{M}_{(j)}\right\rbrace \bigg)\big|}{\sum_{k}\left\{\big|\mathcal{O}_{k}-\mathcal{O}_{\rm av}\big|+\big|\mathcal{T}_{k}\bigg(r_{2},\left \lbrace a_{(j)}, \cos i_{(j)},\dot{M}_{(j)}\right\rbrace \bigg)-\mathcal{O}_{\rm av}\big| \right\}}
\end{align}

\begin{figure*}
\hspace{-2.1cm}
\includegraphics[scale=0.63]{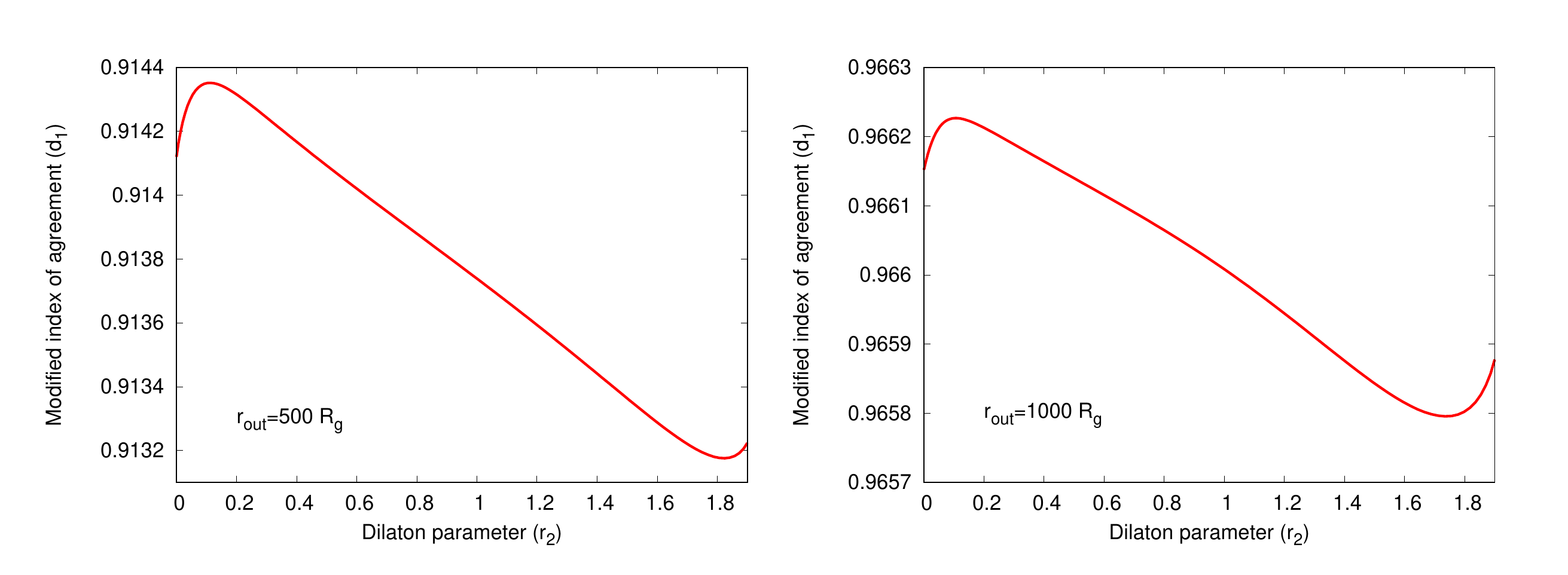}
\caption{The variation of the modified version of the index of agreement $d_1$ with $r_2$, computed with $r_{out}=500r_g$ (left panel) and $r_{out}=1000r_g$ (right panel) is shown in the figure above. Once again the modified index of agreement also maximises for $r_2\sim 0.1$, regardless of the choice of $r_{out}$. See text for more discussions.
}
\label{Fig_6}
\end{figure*}

It is clear from \ref{d} and \ref{d1} that the dilaton parameter which maximizes the index of agreement and its modified form best explain the data. \ref{Fig_5} and \ref{Fig_6} respectively illustrate the variation of $d$ and $d_1$ with $r_2$ for both $r_{out}=500r_g$ and $r_{out}=1000r_g$. We note that both $d$ and $d_1$ maximize when $r_2\sim 0.1$, regardless of the choice of $r_{out}$, in agreement with our earlier results. Also the most favored dilaton parameter $r_2\sim 0.1$, obtained from $E$, $E_1$, $d$ and $d_1$ lies within 68\% confidence level if the $\chi^2$ analysis is considered (see e.g. \ref{Fig_2}), irrespective of the choice of $r_{out}$.   

}
\end{itemize}
By studying the variation of the error estimators with respect to $r_2$ we note that mildly charged axion-dilaton black holes are more favored compared to the Kerr scenario in \gr. This is indicated by all the error estimators since the $\chi^2$ minimizes for $r_2\sim 0.2$ while the other error estimators maximize for $r_2\sim 0.1$. This is further interesting because evaluating the error estimators for different choice of $r_{out}$ keep the result unchanged. Moreover, we plot the confidence intervals ($\Delta \chi^2$ corresponding to a single parameter) about $\chi^2 _{min}$ (\ref{Fig_2}) and note that for $r_{out}=500r_g$,  $r_2=0$ just falls outside 1-$\sigma$ but within 2-$\sigma$ interval (left panel of \ref{Fig_2}), while for $r_{out}=1000r_g$, $r_2=0$ is included in the 1-$\sigma$ interval (right panel of \ref{Fig_2}). \ref{Fig_2} also reveals that if $r_{out}=500r_g$ is considered then $r_2>0.6$ falls outside the 3-$\sigma$ confidence interval while with $r_{out}=1000r_g$, $r_2\lesssim 1.6$ is included within the 3-$\sigma$ confidence level. In any case, our analysis indicates that $r_2>1.6$ is disfavored by optical observations of quasars since it falls outside the 3-$\sigma$ confidence level for both choices of the outer radius. 

It is interesting to compare our findings with \citet{Banerjee:2020ubc} where the authors investigate the effect of the Einstein-Maxwell dilaton axion gravity on the power associated with ballistic jets and the radiative efficiencies derived from the continuum spectrum. The theoretical jet power and the radiative efficiencies derived in the Kerr-Sen background when compared with the available observations of microquasars reveal that \gr\ is more favored compared to the Einstein-Maxwell dilaton-axion scenario.

It is however important to note that there exists several alternative gravity theories whose black hole solutions resemble the Kerr geometry in \gr. Therefore, an observational verification of the Kerr solution cannot distinguish these modified gravity models from \gr \citep{Psaltis:2007cw}. On the other hand, if deviations from \gr\ are detected, then it will require revisiting our understanding of gravity in the high curvature domain. 
Interestingly, a similar analysis with the present quasar sample, when performed with braneworld blackholes (resembling the Kerr-Newman spacetime, although the tidal charge parameter can also accommodate negative values) indicate that quasars with a negative tidal charge (realised in a higher dimensional scenario) are more favored \citep{Banerjee:2017hzw,Banerjee:2019sae}.\\
Since quasars are rotating in nature it is worthwhile to provide an estimate of the spins of the quasars from the above analysis. This is addressed in the next section.

\subsection{Estimating the spins of the quasars}
\label{S4-2}
We have already discussed in the last section that the dilaton parameter which best describes the optical observations of quasars corresponds to $r_2\sim 0.2$ if results based on $\chi^2$ analysis is considered or $r_2\sim 0.1$ if other error estimators are taken into account. In this section we report the most favored values of the spins of the quasars from the same observations. The procedure for extracting the observationally favored spin parameters of the quasar sample, for a given $r_2$, has already been  discussed in \ref{S4}. Here, we consider the magnitude of $r_2\sim 0.2$ corresponding to $\chi^2_{min}$ and report the spins of the quasars which minimize the error between the theoretical and the observed optical luminosities. We also mention the spins of the quasars for $r_2=0$ (general relativity) for comparison with existing results although it is important to note that those spin estimates are highly model dependent \citep{Brenneman:2013oba,Reynolds:2013qqa,Reynolds:2019uxi} 
These are reported in \ref{Table1}.

\begin{table}
\begin{minipage}{60mm}
\small
\hfill{}

\caption{Spin parameters of quasars corresponding to $r_{2}=0.2$ and $r_{2}=0$ (for comparison with GR)}
\label{Table1}
\vspace{-0.36cm}
\begin{center}
\centering
\begin{tabular}{|c|c|c|c|c|c|}

\hline
$\rm Object$ & $\rm log~ m$ &  $\rm log ~L_{obs}$ & $\rm log ~L_{bol}$ & $a_{_{r_{2}=0.2}}$ & $a_{_{r_{2}=0}}$\\
\hline
$\rm 0003+199$ & $\rm 6.88$ &  $\rm 43.91$ & $\rm 45.13 \pm 0.35$ & $\rm  0.9 $ & $\rm 0.99 $\\ \hline
$\rm 0050+124$ & $\rm 6.99$  & $\rm 44.41$ & $\rm 45.12 \pm 0.04$ & $\rm 0.9 $ & $\rm 0.99 $\\ \hline
$\rm 0923+129$ & $\rm 6.82$  & $\rm 43.58$ & $\rm 44.53 \pm 0.15$ & $\rm 0.9 $ & $\rm 0.99 $\\ \hline
$\rm 1011-040$ & $\rm 6.89$ & $\rm 44.08$ & $\rm 45.02 \pm 0.23$ & $\rm 0.9 $ & $\rm 0.99 $\\ \hline
$\rm 1022+519$ & $\rm 6.63$ & $\rm 43.56$ & $\rm 45.10 \pm 0.39$ & $\rm 0.9 $ & $\rm 0.99$\\ \hline
$\rm 1119+120$ & $\rm 7.04$ & $\rm 44.01$ & $\rm 45.18 \pm 0.34$ & $\rm 0.9 $ & $\rm 0.99 $\\ \hline
$\rm 1244+026$ & $\rm 6.15$ & $\rm 43.70$ & $\rm 44.74 \pm 0.22$ & $\rm 0.9 $ & $\rm 0.99 $\\ \hline
$\rm 1404+226$ & $\rm 6.52$ &  $\rm 44.16$ & $\rm 45.21 \pm 0.26$ & $\rm 0.9 $ & $\rm 0.99 $\\ \hline
$\rm 1425+267$ & $\rm 9.53$ &  $\rm 45.55$ & $\rm 46.35 \pm 0.20$ & $\rm -0.1 $ & $\rm 0.7 $\\ \hline
$\rm 1440+356$ & $\rm 7.09$ & $\rm 44.37$ & $\rm 45.62 \pm 0.29$ & $\rm 0.9 $ & $\rm 0.99 $\\ \hline
$\rm 1535+547$ & $\rm 6.78$ & $\rm 43.90$ & $\rm 44.34 \pm 0.02$ & $\rm 0.9 $ & $\rm 0.99 $\\ \hline
$\rm 1545+210$ & $\rm 9.10$ &  $\rm 45.29$ & $\rm 46.14 \pm 0.13$ & $\rm 0.4 $ & $\rm 0.1 $\\ \hline
$\rm 1552+085$ & $\rm 7.17$  & $\rm 44.50$ & $\rm 45.04 \pm 0.01$ & $\rm 0.9 $ & $\rm 0.99 $\\ \hline
$\rm 1613+658$ & $\rm 8.89$  & $\rm 44.75$ & $\rm 45.89 \pm 0.11$ & $\rm 0.2 $ & $\rm 0.3 $\\ \hline
$\rm 1704+608$ & $\rm 9.29$ &  $\rm 45.65$ & $\rm 46.67 \pm 0.21$ & $\rm 0.3 $ & $\rm -0.4 $\\ \hline
$\rm 2308+098$ & $\rm 9.43$ & $\rm 45.62$ & $\rm 46.61 \pm 0.22$ & $\rm -0.4 $ & $\rm 0.95 $\\ \hline
\end{tabular}
\hfill{}
\end{center}
\end{minipage}
\end{table}

We have already discussed in \ref{S4-1} that the theoretically estimated optical luminosity $L_{opt}$ depends both on the radius of the marginally stable circular orbit $r_{ms}$ and the outer extension of the accretion disk $r_{\rm out}$ (\ref{S3-11}). However, emission from the inner radius has a much greater contribution to the total disk luminosity compared to the outer parts of the disk, since the flux from the accretion disk peaks close to the marginally stable circular orbit $r_{ms}$. Therefore, the choice of the outer disk radius $r_{\rm out}$ is not expected to significantly affect our results. Nevertheless we consider both $r_{out}=500r_g$ and $r_{out}=1000r_g$ \citep{Walton:2012aw,Bambi:2011jq} in evaluating the error estimators and confirm that the observationally favored magnitude of $r_2$ is unaffected by the choice of $r_{out}$.

\par
Interestingly, for some of the quasars in our sample, changing $r_{\rm out}$ does affect the best choice of spins corresponding to a given $r_2$. This is because, we allow the accretion rates and the inclination angles to vary (see \ref{S4}). Moreover, apart from the metric parameters, the theoretical optical luminosity is sensitive to $\mathcal{M}$ and $\dot{M}$ which varies between quasars. The ratio $\dot{M}/\mathcal{M}^2$ determines the sharpness of the temperature profile $T(r)$ near $r_{ms}$ (see \ref{S3-8}, \ref{S3-9} and \ref{S3-11}), for a given $r_2$, $a$ and $\cos i$. In the event, $T(r)$ is sharply peaked near $r_{ms}$ the outer disk has negligible contribution to the disk luminosity and the choice of $r_{out}$ does not play a significant role in such cases.
Therefore we report the spins of only those quasars in \ref{Table1} which remain unaltered by variation of $r_{out}$.

We note from \ref{Table1} that most of the quasars are maximally spinning with $a \sim 0.9$ for $r_2\sim 0.2$ while $a \sim 0.99$ when GR is assumed. By using the general relativistic disk reflection model \cite{Ross:2005dm}, \cite{Crummy:2005nj}  
studied the spectra of several quasars reported in \ref{Table1} (PG 0003+199, PG 0050+124, PG 1244+026, PG 1404+226, PG 1440+356) and arrived at a similar conclusion. 
In particular, the spin of PG 0003+199 (also known as Mrk335) is very well constrained, namely, $a \sim 0.89\pm 0.05$ \citep{Keek:2015apa} and 
$a\sim 0.83^{+0.09}_{-0.13}$ \citep{Walton:2012aw} from the reflection spectrum. 
These estimates are very much in agreement with our results based on \gr, although their method of constraining the spin is different from ours. A study of the gravitationally red-shifted iron line of PG 1613+658 (Mrk 876) reveals that the quasar harbors a rotating central black hole which is in agreement with our findings \citep{Bottacini:2014lva}. From the polarimetric observations of AGNs, \cite{Afanasiev:2018dyv} independently estimated the spins of some of the quasars reported in \ref{Table1}. Such an analysis corroborates our spin estimates for PG 0003+199, PG 0050+124, PG 0923+129 and PG 2308+098 although the results for PG 1022+519, PG 1425+267, PG 1545+210, PG 1613+658 and PG 1704+608 show some variations. 
\cite{2017Ap&SS.362..231P} constrained the spin of PG 1704+608 (3C 351) to $a<0.998$, consistent with our results. It turns out that that rapidly rotating prograde systems are radio-quiet \citep{Kellermann:1989tq,Barvainis:2004wr,Sikora:2006xz,Villforth:2009eq}, which is in accordance with our findings.
\\

\section{Summary and concluding remarks}
\label{S5}
In this work we attempt to discern the signatures of Einstein-Maxwell dilaton-axion (EMDA) gravity from the quasar continuum spectrum, which is believed to be an important astrophysical site to examine the nature of gravitational interaction in the high curvature regime. EMDA gravity arises in the low energy effective action of the heterotic string theory and bears the coupling of the scalar dilaton and the pseudo-scalar axion to the metric and the Maxwell field. The dilaton and the axion fields are inherited in the action from string compactifications and exhibit interesting implications in  inflationary cosmology and the late time acceleration of the universe. Therefore, it is worthwhile to search for the imprints of these fields in the available astrophysical observations since this provides a testbed for string theory.

The presence of dilaton and axion in the theory results in substantial modifications of the gravitational field equations compared to \gr. The stationary and axi-symmetric black hole solution of these equations leads to the Kerr-Sen metric which carries a dilaton charge with the axionic field imparting angular momentum to the black hole. The presence of the Maxwell field makes the Kerr-Sen black hole electrically charged and renders non-trivial field strengths to the axion and the dilaton. The electric charge of the black hole, however, stems from the axion-photon coupling and not the in-falling charged particles. In the absence of the Maxwell field the effect of dilaton and axion vanish and the Kerr-Sen metric reduces to the Kerr metric in \gr.   

The observational signatures of the Kerr-Sen spacetime has been explored in the context of strong gravitational lensing and black hole shadows \citep{Gyulchev:2006zg,An:2017hby,Younsi:2016azx,Hioki:2008zw}, although these works establish no constrains on the dilaton parameter based on the available observations. An attempt has been recently made \citep{Heydari-fard:2020syf} to investigate the impact of the dilaton-axion black holes on the black hole continuum spectrum, although there, the authors have not compared the theoretical spectra with the observations.

In this work, on the other hand, we compute the theoretical estimate of optical luminosity for a sample of eighty Palomar Green quasars using the thin accretion disk model proposed by Novikov \& Thorne. 
These are then compared with the corresponding optical observations of quasars to obtain an estimate of the observationally favored dilaton parameter and the angular momentum of the quasars. Our study brings out that the dilaton parameter most favored by observations corresponds to $r_2\sim 0.2$ and if $r_{out}=500r_g$ is considered then
$0.05 \lesssim r_2\lesssim 0.3$ lies within 1-$\sigma$ confidence interval. Extending $r_{out}$ to $1000r_g$ brings $0\lesssim r_2\lesssim 0.5$ within the 1-$\sigma$ confidence level.  
Also $r_2>1.6$ seems to be disfavored by quasar optical data since it lies outside 3-$\sigma$ confidence level irrespective of the choice of $r_{out}$. We further note that in \cite{Banerjee:2020ubc}, the authors constrain the dilaton parameter by investigating the observed jet power and the radiative efficiencies of the microquasars and report that $r_2\sim 0$ is favored by such observations.
Also it is worthwhile to mention that an observational verification of the Kerr solution ($r_2= 0$) does not confirm \gr\ with certainty, since apart from \gr, the Kerr metric represents the black hole solution for several other alternate gravity models \citep{Psaltis:2007cw}.

The fact that strong dilaton charges are disfavored by observations indicates a minuscle field strength for axion whose suppression has been observed in several other physical scenarios, e.g. in the inflationary era induced by higher curvature gravity \cite{Elizalde:2018rmz,Elizalde:2018now} and higher dimensions \cite{Paul:2018jpq}, in the warped braneworld scenario \cite{Randall:1999ee} with bulk Kalb-Ramond fields \cite{Mukhopadhyaya:2001fc,Mukhopadhyaya:2002jn} and the related stabilization of the modulus \cite{Das:2014asa} and so on.

The present analysis also allows us to constrain the spins of the quasars which are mostly in agreement with the previous estimates. However, there are limitations associated with spin measurements. This is because the spectral energy distribution (SED) of the quasars consists of emission from multiple components, e.g. the accretion disk, the corona, the jet and the dusty torus which are not always easy to observe and model \citep{Brenneman:2013oba}. Discerning the effect of each of these components from the SED is often very challenging which limits accurate determinination of their mass, spin, distance and inclination. As a result the spin of the same quasar estimated by different methods often leads to inconsistent results \citep{Brenneman:2013oba,Reynolds:2013qqa,Steiner:2012vq,Gou:2009ks,Reynolds:2019uxi}.

We further note that the background metric affects the emission from only those components which reside close to the horizon. We are therefore interested in modelling the continuum spectrum from the accretion disk since its inner edge approaches the vicinity of the horizon. The continuum spectrum however, is not only sensitive to the background spacetime but also on the properties of the accretion flow. In the present work the spectrum is computed using the thin-disk approximation which does not take into account the presence of outflows or the radial velocity of the accretion flow. A more comprehensive modelling of the disk would therefore impose stronger constraints on the background metric. At present, these issues are addressed by considering several phenomenological models which is beyond the scope of this work.

Apart from the continuum spectrum, there exists other astrophysical observations e.g., the quasi-periodic oscillations observed in the power spectrum of black holes \citep{Maselli:2014fca,Pappas:2012nt}, the broadened and skewed iron K-$\alpha$ line in the reflection spectrum of black holes \citep{Bambi:2016sac,Ni:2016uik}, and the black hole shadow \citep{Akiyama:2019brx,Akiyama:2019fyp,Akiyama:2019eap}, which can be used to further establish or falsify our present findings. We leave this study for a future work which will be reported elsewhere.

\section{Data Availability}
The data underlying this article are taken from \cite{Davis:2010uq}
cited in the text.


\section{Acknowledgements}
The research of SSG is partially supported by the Science and Engineering Research Board-Extra Mural Research Grant No. (EMR/2017/001372), Government of India.

\small{\bibliographystyle{mnras}}
\bibliography{accretion,KN-ED,EMDA-Jet}


\end{document}